\documentclass[english]{article}
\usepackage[T1]{fontenc}
\usepackage[a4paper]{geometry}
\usepackage{url}
\usepackage{amsmath}
\usepackage{amsthm}
\usepackage{amssymb}
\usepackage{graphicx}
\usepackage{xargs}[2008/03/08]

\makeatletter

\usepackage{tikz}
\usetikzlibrary{calc}

\makeatother

\usepackage{babel}
\usepackage[style=authoryear,backend=bibtex]{biblatex}
\begin{document}
\title{A Mixed Integer Least-Squares Formulation of the GNSS Snapshot Positioning
Problem}
\author{Eyal Waserman and Sivan Toledo\\
Blavatnik School of Computer Science, Tel-Aviv University}
\maketitle
\begin{abstract}
This paper presents a formulation of \emph{Snapshot Positioning} as
a mixed-integer least-squares problem. In snapshot positioning, one
estimates a position from code-phase (and possibly Doppler-shift)
observations of Global Navigation Satellite Systems (GNSS) signals
without knowing the time of departure (timestamp) of the codes. Solving
the problem allows a receiver to determine a fix from short radio-frequency
snapshots missing the timestamp information embedded in the GNSS data
stream. This is used to reduce the time to first fix in some receivers,
and it is used in certain wildlife trackers. This paper presents two
new formulations of the problem and an algorithm that solves the resulting
mixed-integer least-squares problems. We also show that the new formulations
can produce fixes even with huge initial errors, much larger than
permitted in Van Diggelen's widely-cited \emph{coarse-time navigation
}method.
\end{abstract}

\global\long\def\linearestimator{F}%

\global\long\def\modelfun{M}%

\global\long\def\penaltyfun{\phi}%

\global\long\def\objectivefun{\phi}%

\global\long\def\grad{\nabla}%

\newcommandx\hessian[1][usedefault, addprefix=\global, 1=]{\nabla_{#1}^{2}}%

\global\long\def\jacobian{\mathrm{J}}%

\global\long\def\exact#1{\mathring{#1}}%

\global\long\def\estimate#1{\hat{#1}}%

\global\long\def\controlpoint{\rho}%

\global\long\def\location{\ell}%

\global\long\def\noise{\epsilon}%

\global\long\def\observations{b}%

\global\long\def\expectation{\mathrm{E}}%

\global\long\def\iu{\mathbf{i}}%

\global\long\def\vecone{\mathbf{1}}%

\global\long\def\veczero{\mathbf{0}}%

\global\long\def\cov{\text{cov}}%

\global\long\def\var{\text{var}}%

\global\long\def\fim{\mathbb{I}}%

\global\long\def\duration{\vartheta}%

\global\long\def\initialphase{\varphi}%

\global\long\def\xcorr{\operatorname{xcorr}}%

\global\long\def\diag{\operatorname{diag}}%

\global\long\def\rank{\operatorname{rank}}%

\global\long\def\erf{\operatorname{erf}}%

\global\long\def\real{\text{Re}}%

\global\long\def\imag{\text{Im}}%

\global\long\def\square{\text{\ensuremath{\blacksquare}}}%


\section{\label{sec:Introduction}Introduction}

\noindent 

The fundamental observation equation in Global Navigation Satellite
Systems (GNSS) is
\[
t_{i}-t_{D,i}=\frac{1}{c}\left\Vert \exact{\location}-\controlpoint_{i}\left(t_{D,i}\right)\right\Vert _{2}+\exact b+\delta_{i}+\noise_{i}\;,
\]
where $t_{i}$ is the observed (estimated) time of arrival of a code
from satellite $i$, $t_{D,i}$ is the time of departure of the signal,
$c$ is the speed of light, $\exact{\location}$ is the location of
the receiver, $\controlpoint_{i}(t_{D,i})$ is the location of the
satellite at the time of transmission, $\exact b$ is the offset in
time-of-arrival observation caused by the inaccurate clock at the
receiver and by delays in the analog RF chain (e.g., in cables), $\delta_{i}$
represents atmospheric delays and the satellite's clock error, and
$\noise_{i}$ is an error or noise term that accounts for both physical
noise and for unmodeled effects. Normally, the GNSS solver estimates
$\exact{\location}$ and $\exact b$ by minimizing the norm of the
error vector $\noise$; it does not know $\noise_{i}$ and does not
attempt to estimate it. The quantities $t_{D,i}$ and $\controlpoint_{i}(t_{D,i})$
are usually known; $t_{D,i}$ is known because the satellite timestamps
its transmission, and $\controlpoint_{i}(t_{D,i})$ is known because
the satellite transmits the parameters that define its orbit, called
the \emph{ephemeris}. The ephemeris can also be downloaded from the
Internet. 

Decoding $t_{D,i}$ takes a significant amount of time, in GPS up
to 6 seconds under good SNR conditions and longer in low-SNR conditions.
GNSS receivers that need to log locations by observing the RF signals
for short periods cannot decode $t_{D,i}$. Examples for such applications
include tracking marine animals like sea turtles, which surface briefly
and then submerge again. It turns out that techniques that are collectively
called \emph{snapshot positioning} or \emph{coarse-time navigation}
can estimate $\exact{\location}$ and $\exact b$ when $t_{D,i}$
is not known. These techniques can also reduce the time to a first
fix when a receiver is turned on.

Snapshot receivers sample the incoming GNSS RF signals for a short
period, called a \emph{snapshot}. Usually (but not always), the RF
samples are correlated with replicas of the codes transmitted by the
satellites, therefore determining $t_{i}$ for the subset of visible
satellites. The correlation (and Doppler search) are sometimes performed
on the receiver, which then stores or transmits the $t_{i}$s. This
appears to be the case for a proprietary technology called Fastloc,
which is used primarily to track marine animals~\parencite{doi:10.1111/2041-210X.12286,Tomkiewicz,WITT2010571}.
In other cases~\parencite{Robin2,Robin1,YovelCellGuide2,10.1145/3301293.3302367,HartenEtAlScienceJuly2020,GPSCloudOffloading,10.1145/2030112.2030158},
the logger records the raw RF samples and correlation is performed
after the data is uploaded to a computer.

Techniques for estimating $\exact{\location}$ when the $t_{D,i}$s
are not known date back to a 1995 paper by Peterson et al~\parencite{GPSForUrbanCanyon}.
They proposed to view $t_{D,i}$ as a function of both $t_{i}$ and
a coarse clock-error unknown that they call \emph{coarse time}, which
in principle is identical to $\exact b$, but is modeled by a separate
variable. They then show that it is usually possible to estimate $\exact{\location}$,
$\exact b$, and the coarse time from five or more $t_{i}$s. This
method does not always resolve the $t_{D,i}$ correctly. Lannelongue
and Pablos~\parencite{CTN1998} and Van Diggelen~\parencite{IEEEexample:uspat,VanDiggelenAGPS}
proposed methods that appear to always resolve the $t_{D,i}$ correctly
when the initial estimate of $\exact{\location}$ and $\exact b$
are within some limits (adding up to about 150~km). Muthuraman, Brown,
and Chansarkar~\parencite{CTNEquiv} showed that the two methods are equivalent
in the sense that they usually produce the same estimates. However,
the method of Lannelongue and Pablos is an iterative search procedure,
while Van Diggelen's is a rounding procedure that is more computationally
efficient, so Van Diggelen's method became much more widely used and
widely cited. Van Diggelen also showed how to use an iterative procedure
over a  of possible positions when the initial estimate of $\exact{\location}$
and $\exact b$ is outside the 150km limit.

All three methods use a system of linearized equations with five scalar
unknowns (not the usual 4), which are corrections to the coordinates
of $\exact{\location}$, the offset $\exact b$ (which he refers to
as the common bias), and the coarse time unknown, usually denoted
by $tc$ (we denote it in this paper by the single letter $s$). 

Van Diggelen's algorithm was used and cited in numerous subsequent
papers, all of which repeat his presentation without adding explanations.
Liu et al~\parencite{GPSCloudOffloading}, Ramos et al~\parencite{10.1145/2030112.2030158},
and Wang et al~\parencite{TerminalPositioningGPS} describe snapshot GPS
loggers whose recordings are processed using the algorithm. Badia-Sol\'{e}
and Iacobescu Ioan~\parencite{GPSSnapshotTechniquesReport} report on
the performance of the method. Othieno and Gleason~\parencite{Othieno2012CombinedDT},
Chen at al~\parencite{CTNWithDoppler}, and Fern\'{a}ndez-Hern\'{a}ndez
and Borre~\parencite{CTNWithoutInitialInfo} show how to use Doppler measurements
to obtain an initial estimate that satisfies the requirements of Van
Diggelen's algorithm. Yoo et al~\parencite{CTNImproved} propose a technique
that replaces the estimation of the coarse-time parameter by a one-dimensional
 search, which allows them to estimate $\exact{\location}$ using
observations from  four satellites, but at considerable computational
expense.

Bissig et al~\parencite{OneMillisecond} use a \emph{direct position determination}~\parencite{WeissDPD}
approach to snapshot positioning. They quantize the four unknowns
$\exact{\location}$ and $\exact b$ and maximize the likelihood of
the received snapshot over this four-dimensional lattice. To make
the search efficient, they use a branch and bound approach that prunes
sets of unlikely solutions. It appears that this approach allows them
to estimate locations using very short snapshots, but at the cost
of fairly low accuracy and relatively long running times, compared
to methods that estimate the $t_{i}$s first.

In this paper we derive the observation equations that underlie the
methods of Peterson et al.~\parencite{GPSForUrbanCanyon}, Lannelongue
and Pablos~\parencite{CTN1998} and Van Diggelen~\parencite{IEEEexample:uspat,VanDiggelenAGPS}.
These authors  show the correction equations, not the observation
equations whose Jacobian constitutes the correction equations. The
formulation of the observation equations, which constitute a mixed-integer
least-squares problem, allows us to apply a new type of algorithm
to estimate the integer unknowns. A mixed-integer least-squares problem
is an optimization problem with a least-squares objective function
and both real (continuous) unknowns and integer unknowns. More specifically,
we regularize the mixed-integer problem using either a priori estimates
of $\exact{\location}$ and $\exact b$ or Doppler-shift observations.
Our approach is inspired by the \emph{real-time kinematic} (RTK) method,
which resolves a position from both code-phase and carrier-phase GNSS
observations~\parencite{GNSSHandbookIntegerAmbiguity}; carrier-phase
constraints have integer ambiguities that must be resolved.

Our experimental results using real-world data demonstrate that our
new algorithms can resolve locations with much larger initial location
and time errors than the method of Van Diggelen. Van Diggelen's non-iterative
method only works when the initial estimate is up to about 150~km
(or equivalent combinations), whereas our mixed-integer least-squares
solver works with initial errors of up 150~s and 200~km. When using
Doppler-shift regularization, our method works even with initial errors
of 180~s and arbitrarily large initial position errors (on Earth);
if the initial position error is small, the method tolerates initial
time errors of up to 5000~s.

Our implementation of the new methods and the code that we used to
evaluate them are publicly available\footnote{\url{https://github.com/eyalw711/snapshot-positioning}}. 

The rest of this paper is organized as follows. Section~\ref{sec:models-and-algorithms}
presents the observation equations for the snapshot-positioning GNSS
problem. Section~\ref{subsec:Shadowing} explains how to incorporate
the so-called \emph{coarse-time} parameter into the observation equations
and how Van Diggelen's method exploits it. Section~\ref{subsec:apriori-regularization}
presents our first regularized formulation, which uses the initial
guess to regularize the mixed-integer least-squares problem. Section~\ref{subsec:Doppler-Regularization}
presents the Doppler-regularized formulation. Our experimental results
are presented in Section~\ref{sec:Implementation-and-Evaluation}.
Section~\ref{sec:Conclusions-and-Discussion} discusses our conclusions
from this research.

\section{\label{sec:models-and-algorithms}The Snapshot-Positioning Problem:
A GNSS Model with Whole-Millisecond Ambiguities}

We begin by showing that when departure times are not known, the observation
equations that relate the arrival times of GNSS codes to the unknown
position of the receiver contain integer ambiguities.

We denote by $t_{D,i}$ the time of departure of a code from satellite
$i$, and we assume that $t_{D,i}$ represents a whole millisecond
(in the time base of the GPS system). We denote by $\exact t_{i}$
the time of arrival of that code at the antenna of the receiver. We
assume that the receiver estimates the arrival time of a that code
as $t_{i}=\exact t_{i}+\exact b+\epsilon_{i}$, where $\exact b$
represents the bias that is caused by the inaccurate clock of the
receiver and by delays that the signal experiences in the path from
the antenna to the ADC and $\epsilon_{i}$ is the arrival-time estimation
error. The bias $\exact b$ is time dependent, because of drift in
the receiver's clock, but over short observation periods this dependence
is negligible, so we ignore it.

The time of arrival is governed by the equation
\[
\exact t_{i}-t_{D,i}=\frac{1}{c}\left\Vert \exact{\location}-\controlpoint_{i}\left(t_{D,i}\right)\right\Vert _{2}+\delta_{i}\;,
\]
where $c$ is the speed of light, $\exact{\location}$ is the location
of the receiver, $\controlpoint_{i}$ is the location of the satellite
(which is a function of time, since the satellites are not stationary
relative to Earth observers), and $\delta_{i}$ represents the inaccuracy
of the satellite's clock and atmospheric delays. We assume that $\delta_{i}$
can be modeled, for example using models of ionospheric and tropospheric
delays (dual frequency receivers can estimate the ionospheric delay,
but we assume a single-frequency receiver). Setting $\delta_{i}$
to the satellite's clock error correction from the ephemeris induces
a location error of about 30m due to the atmospheric delays~\parencite{StrangBorreGPS2}.

A receiver that decodes the timestamp embedded in the GPS data stream
can determine $t_{D,i}$, which leads to the following equation
\[
t_{i}-t_{D,i}=\frac{1}{c}\left\Vert \exact{\location}-\controlpoint_{i}\left(t_{D,i}\right)\right\Vert _{2}+\exact b+\delta_{i}+\noise_{i}\;,
\]
the conventional GNSS code-observation equation, in which the $4$
unknown parameters are $\exact b$ and the coordinates of $\exact{\location}$
(we assume that $\delta_{i}$ is modeled, possibly trivially $\delta_{i}=0$,
but not estimated). To simplify the notation, we ignore $\delta_{i}$
for now and write 
\[
t_{i}-t_{D,i}=\frac{1}{c}\left\Vert \exact{\location}-\controlpoint_{i}\left(t_{D,i}\right)\right\Vert _{2}+\exact b+\noise_{i}\;.
\]
We use observations from all the satellites such that all the $t_{i}$s
lie between two consecutive whole multiples of $t_{\text{code}}$
(in GPS, two round milliseconds in the local clock). This allows us
to express 
\[
t_{i}=\left(N+\varphi_{i}\right)t_{\text{code}}
\]
with a common and easily computable $N=\left\lfloor t_{i}/t_{\text{code}}\right\rfloor $
and for $\varphi_{i}\in[0,1)$. We denote $N_{i}=t_{D,i}/t_{\text{code}}$
and write 
\[
\left(N-N_{i}+\varphi_{i}\right)t_{\text{code}}=\frac{1}{c}\left\Vert \exact{\location}-\controlpoint_{i}\left(t_{D,i}\right)\right\Vert _{2}+\exact b+\noise_{i}\;.
\]
Since GNSS codes are aligned with $t_{\text{code}},$ $N_{i}\in\mathbb{Z}$.
We denote $n_{i}=N-N_{i}\in\mathbb{Z}$, so 
\[
\left(n_{i}+\varphi_{i}\right)t_{\text{code}}=\frac{1}{c}\left\Vert \exact{\location}-\controlpoint_{i}\left(t_{D,i}\right)\right\Vert _{2}+\exact b+\noise_{i}
\]
or
\[
\varphi_{i}t_{\text{code}}=\frac{1}{c}\left\Vert \exact{\location}-\controlpoint_{i}\left(t_{D,i}\right)\right\Vert _{2}-n_{i}t_{\text{code}}+\exact b+\noise_{i}\;.
\]
We now face two challenges. One is that we have $4+m$ unknown parameters:
three location coordinates, $\exact b$, and the $n_{i}$s, but only
$m$ constraints. We clearly need more constraints so that we can
resolve the $n_{i}$s. The other is that we have a set of nonlinear
constraints with continuous real unknowns, the location and $\exact b$,
and with integer unknowns, the $n_{i}$. The strategy, as in other
cases with this structure, is to first linearize the non-linear term,
then to resolve the integer parameters, and to then substitute them
and to solve the continuous least-squares problem (either the linearized
system or the original non-linear system). We cannot linearize the
non-linear term $\|\exact{\location}-\controlpoint_{i}(t_{D,i})\|_{2}/c$
using a Taylor series because it is a function of both real unknowns
and  of the integer unknowns $n_{i}$. We cannot differentiate this
term with respect to the integer $n_{i}$. 

To address this difficulty, we approximate $t_{D,i}$ by approximating
the range (distance) term in the equation
\[
t_{D,i}=t_{i}-\frac{1}{c}\left\Vert \exact{\location}-\controlpoint_{i}\left(t_{D,i}\right)\right\Vert _{2}-\exact b-\noise_{i}\;.
\]
For now, we denote the approximation of the propagation delay by 
\[
d_{i}\approx\frac{1}{c}\left\Vert \exact{\location}-\controlpoint_{i}\left(t_{D,i}\right)\right\Vert _{2}+\noise_{i}
\]
so
\[
\estimate t_{D,i}=t_{i}-d_{i}-\exact b\;.
\]
There are several ways to set $d_{i}$, depending on our prior knowledge
of $\exact{\location}$ and $\exact b$. One option in the GPS system
is to set it to about $76.5\text{ms}$;this limits the error in $\estimate t_{D,i}$
to about $12.5\text{ms}$ for any Earth observer, and the error
\[
\left\Vert \exact{\location}-\controlpoint_{i}\left(\estimate t_{D,i}\right)\right\Vert _{2}-\left\Vert \exact{\location}-\controlpoint_{i}\left(t_{D,i}\right)\right\Vert _{2}
\]
to about $10\text{m}$~\parencite{VanDiggelenAGPS}. We substitute $\controlpoint_{i}(\estimate t_{D,i})=\controlpoint_{i}(t_{i}-d_{i}-\exact b)$
for $\controlpoint_{i}(t_{D,i})$,
\begin{equation}
\varphi_{i}t_{\text{code}}=\frac{1}{c}\left\Vert \exact{\location}-\controlpoint_{i}\left(t_{i}-d_{i}-\exact b\right)\right\Vert _{2}-n_{i}t_{\text{code}}+\exact b+\noise_{i}^{(D)}\;.\label{eq:nonlin-inexact-D}
\end{equation}
The superscript $(D)$ on the error term indicates that the error
term now represents not only the arrival-time estimation error, but
also the error induced by the inexact departure time. 

We linearize around an a priori solution $\bar{\location}$ and $\bar{b}$
(usually $\bar{b}=0$, otherwise we can simply shift the $t_{i}$s),
\begin{eqnarray}
\varphi_{i}t_{\text{code}} & = & \frac{1}{c}\left\Vert \bar{\location}-\controlpoint_{i}\left(t_{i}-d_{i}-\bar{b}\right)\right\Vert _{2}+\frac{1}{c}\jacobian_{i,:}\begin{bmatrix}\exact{\location}-\bar{\ell}\\
\exact b-\bar{b}
\end{bmatrix}-n_{i}t_{\text{code}}+\exact b+\noise_{i}^{(D,L)}\nonumber \\
 & = & \frac{1}{c}\left\Vert \bar{\location}-\controlpoint_{i}\left(t_{i}-d_{i}-\bar{b}\right)\right\Vert _{2}+\frac{1}{c}\jacobian_{i,:}\begin{bmatrix}\exact{\location}-\bar{\ell}\\
\exact b-\bar{b}
\end{bmatrix}-n_{i}t_{\text{code}}+\left(\exact b-\bar{b}\right)+\bar{b}+\noise_{i}^{(D,L)}\;,\label{eq:linearized-inexact-D-L}
\end{eqnarray}
where $\jacobian$ is the Jacobian of the Euclidean distances with
respect to both the location of the receiver and to the bias, with
the derivatives evaluated at $\bar{\location}$ and at $t_{i}-d_{i}-\bar{b}$.
The superscript $(D,L)$ on the error term indicates that it includes
now also the linearization error.

There are now several ways to resolve the $n_{i}$. 

\section{\label{subsec:Shadowing}Shadowing}

Peterson et al.~\parencite{GPSForUrbanCanyon} introduced a somewhat surprising
modeling technique, which we refer to as \emph{shadowing}. The idea
is to replace the unknown $b$ by two separate unknowns that represent
essentially the same quantity, the original $b$ and a shadow $s$.
In principle, they should obey the equation $b=s$, but the model
treats $s$ as a free parameter; the constraint $b=s$ is dropped.
In the literature, $s$ is called the \emph{coarse-time} parameter
(and is often represented by $tc$ or $t_{c}$). We express this technique
by splitting $b$ and $s$:
\begin{eqnarray*}
\varphi_{i}t_{\text{code}} & = & \frac{1}{c}\left\Vert \bar{\location}-\controlpoint_{i}\left(t_{i}-d_{i}-\bar{b}\right)\right\Vert _{2}+\frac{1}{c}\jacobian_{i,:}\begin{bmatrix}\exact{\location}-\bar{\ell}\\
\exact b-\bar{b}
\end{bmatrix}-n_{i}t_{\text{code}}+\exact b+\noise_{i}^{(D,L)}\\
 & = & \frac{1}{c}\left\Vert \bar{\location}-\controlpoint_{i}\left(t_{i}-d_{i}-\bar{b}\right)\right\Vert _{2}+\frac{1}{c}\jacobian_{i,:}\begin{bmatrix}\exact{\location}-\bar{\ell}\\
s-\bar{b}
\end{bmatrix}-n_{i}t_{\text{code}}+\exact b+\noise_{i}^{(D,L)}\;.
\end{eqnarray*}
We now have five unknowns, not four. 

As far as we can tell, there is no clear explanation in the literature
as to the benefits of shadowing. One way to justify the technique
is to observe that Equation~(\ref{eq:linearized-inexact-D-L}) is
very sensitive to small (nanosecond scale) perturbations in the additive
$\exact b$, but it is not highly sensitive to the $\exact b$ (now
$s$) that we multiply by 
\begin{equation}
\jacobian_{i,4}=\frac{\partial}{\partial t_{D,i}}\left\Vert \bar{\location}-\controlpoint_{i}\left(t_{i}-d_{i}-\bar{b}\right)\right\Vert _{2}\;.\label{eq:range-rate-multiplier}
\end{equation}
For example, in GPS the derivative is bounded by about $800\text{m/s}$
for any $\bar{\location}$ on Earth~\parencite{VanDiggelenAGPS}, so $\jacobian_{i,4}/c<3\times10^{-6}$
(versus $1$ for the additive $\exact b$).  Therefore, the dependence
of the residual (the vector of $\noise_{i}^{(D,L)}$'s for a given
setting of the unknown parameters) on $\exact b$ in Equation~(\ref{eq:linearized-inexact-D-L})
is highly non-convex. There are many different values of $\exact b$
that are almost equally good, a millisecond apart, with each of these
nearly-optimal hypotheses being locally well defined; if we increase
$\exact b$ by one millisecond and also add $1$ to each $n_{i}$,
the residual changes very little, because $\jacobian{}_{i,4}$ is
so small.

Shadowing turns this non convexity into explicit rank deficiency,
which is easier to deal with. With one instance of $\exact b$ replaced
by the shadow $s$, the constraints no longer uniquely define $\exact b$,
only up to a multiple of $t_{\text{code}}$. For any hypothetical
solution $\location,s,b,n$, the solution $\location,s,b+kt_{\text{code}},n+k$
gives exactly the same residual. We perform a change of variables,
replacing the partial sum $-n_{i}t_{\text{code}}+\exact b$ by $-\nu_{i}t_{\text{code}}+\beta$,
where $-\nu_{i}=-n_{i}+\left\lfloor \exact b/t_{\text{code}}\right\rfloor $
and $\beta=\exact b-\left\lfloor b/t_{\text{code}}\right\rfloor t_{\text{code}}$,
so $\beta\in[0,t_{\text{code}})$:
\begin{eqnarray}
\varphi_{i}t_{\text{code}} & = & \frac{1}{c}\left\Vert \bar{\location}-\controlpoint_{i}\left(t_{i}-d_{i}-\bar{b}\right)\right\Vert _{2}+\frac{1}{c}\jacobian_{i,:}\begin{bmatrix}\exact{\location}-\bar{\ell}\\
s-\bar{b}
\end{bmatrix}-\nu_{i}t_{\text{code}}+\beta+\noise_{i}^{(D,L)}\;.\label{eq:shadow-DL}\\
\beta & \in & [0,t_{\text{code}})\;.\nonumber 
\end{eqnarray}

\subsection{Resolving the Integer Ambiguities: Van Diggelen's Method}

Van Diggelen's method exploits the fact that the $\nu_{i}$'s are
very insensitive to $\exact{\ell}$ and to $s$. It therefore sets
$\exact{\location}=\bar{\ell}$ and $s=\bar{b}$, truncating the Jacobian
term from Equation~(\ref{eq:shadow-DL}):
\begin{eqnarray}
\varphi_{i}t_{\text{code}} & = & \frac{1}{c}\left\Vert \bar{\location}-\controlpoint_{i}\left(t_{i}-d_{i}-\bar{b}\right)\right\Vert _{2}-\nu_{i}t_{\text{code}}+\beta+\noise_{i}^{(D,L,A)}\label{eq:shadow-DLA}\\
\beta & \in & [0,t_{\text{code}})\;.\nonumber 
\end{eqnarray}
The new subscript $(D,L,A)$ indicates that the error term now compensates
also for the use of the a priori estimates $\bar{b}$ and $\bar{\location}$
for $s$ and $\exact{\location}$. 

Van Diggelen uses these constraints to set the $\nu_{i}$'s in a particular
way. The method selects one index $j$ that is used to set $\nu_{j}$
and $\beta$ and then resolves all the other $\nu_{i}$s so they are
consistent with this $\beta$. That is, he assumes that $\noise_{j}^{(D,L,A)}=0$
so\\
\begin{eqnarray*}
\nu_{j} & = & \left\lceil \frac{\frac{1}{c}\left\Vert \bar{\location}-\controlpoint_{j}\left(t_{j}-d_{j}-\bar{b}\right)\right\Vert -\varphi_{j}t_{\text{code}}}{t_{\text{code}}}\right\rceil \\
\beta & = & \text{\ensuremath{\left(\nu_{j}+\varphi_{j}\right)}}t_{\text{code}}-\frac{1}{c}\left\Vert \bar{\location}-\controlpoint_{j}\left(t_{j}-d_{j}-\bar{b}\right)\right\Vert _{2}\;.
\end{eqnarray*}
The method now substitutes this $\beta$ in all the other constraints
and assigns the other $\nu_{i}$s by setting $\noise_{i}^{(D,L,A)}=0$
and rounding,
\begin{equation}
\nu_{i}=\left\lfloor \frac{\frac{1}{c}\left\Vert \bar{\location}-\controlpoint_{i}\left(t_{i}-d_{i}-\bar{b}\right)\right\Vert -\varphi_{i}t_{\text{code}}+\beta}{t_{\text{code}}}\right\rceil \;.\label{eq:round_nis}
\end{equation}
When $\|\exact{\ell}-\bar{\ell}\|$ and $|s-\bar{b}|$ are small enough,
this gives a set of $\nu_{i}$s that are correct in the sense that
they all differ from the correct $\nu_{i}$s by the same integer. 

Van Diggelen chooses $j$ in a particular way: he chooses the $j$
that minimizes the magnitude of~(\ref{eq:range-rate-multiplier}),
which corresponds to the satellite closest to the zenith of $\bar{\location}$
at $t_{j}-\bar{b}$. In our notation, Van Diggelen's justification
for this choice is as follows. He searches for a $j$ for which Equation~(\ref{eq:shadow-DLA})
approximates well Equation~(\ref{eq:shadow-DL}). The difference
between the two is
\[
\frac{1}{c}\jacobian_{j,:}\begin{bmatrix}\exact{\location}-\bar{\ell}\\
s-\bar{b}
\end{bmatrix}\;.
\]
For each satellite, $\jacobian_{i,1:3}$ is the negation of the so-called
line-of-sight vector $e_{i}^{T}$, which is the normalized direction
from the satellite to the receiver; element $\jacobian_{i,4}$ is
the range-rate. Van Diggelen's choice of $j$ leads to a row of $\jacobian$
in which the first three elements are almost orthogonal to $\exact{\location}-\bar{\ell}$
and in which the fourth element, the range rate, is small. This leads
to an estimated $\beta$ that is relatively accurate, which helps
resolve the correct $\nu_{i}$'s.

Van Diggelen also shows that if we resolve the $\nu_{i}$'s by setting
each $\noise_{i}^{(D,L,A)}=0$ \emph{separately}, then the resolved
$\beta$'s might be close to $0$ in one equation and close to $t_{\text{code}}$
in another; this leads to inconsistent $\nu_{i}$'s and to a huge
position error.

\subsection{Final Resolution of the Receiver's Location}

Van Diggelen's method resolves the integer $\nu_{i}$'s in Equation~(\ref{eq:shadow-DLA}).
Now we need to resolve the continuous unknowns. We do so using Gauss-Newton
iterations on Equation~(\ref{eq:shadow-DL}), iterating on $\delta_{\location}=\exact{\location}-\bar{\location}$,
$\delta_{s}=s-\bar{b}$, and $\beta$ but keeping $\nu$ fixed. We
start with $\delta_{\location}$, $\delta_{s}$, and $\beta$ set
to zero.

In every iteration, we use the current iterates to produce estimates
of the location and bias,
\begin{eqnarray*}
\estimate{\location} & = & \bar{\location}+\delta_{\location}\\
\estimate b & = & \bar{b}+\delta_{s}\;.
\end{eqnarray*}
We use them to improve the estimate of the ranges $d_{i}$, setting
\[
\estimate d_{i}=\left\Vert \estimate{\location}-\controlpoint_{i}\left(\estimate t_{D,i}\right)\right\Vert _{2}\;.
\]
This allows us to reduce the errors in Equation~(\ref{eq:nonlin-inexact-D}),
\begin{eqnarray*}
\varphi_{i}t_{\text{code}} & = & \frac{1}{c}\left\Vert \exact{\location}-\controlpoint_{i}\left(t_{i}-\estimate d_{i}-\exact b\right)\right\Vert _{2}-n_{i}t_{\text{code}}+\exact b+\noise_{i}^{(\estimate D)}\\
 & = & \frac{1}{c}\left\Vert \exact{\location}-\controlpoint_{i}\left(t_{i}-\estimate d_{i}-\exact b\right)\right\Vert _{2}-\nu_{i}t_{\text{code}}+\beta+\noise_{i}^{(\estimate D)}
\end{eqnarray*}
(the second line holds because $n_{i}t_{\text{code}}+\exact b=\nu_{i}t_{\text{code}}+\beta$,
by definition). We again linearize this and solve the constraints
\begin{equation}
\varphi_{i}t_{\text{code}}=\frac{1}{c}\left\Vert \estimate{\location}-\controlpoint_{i}\left(t_{i}-\estimate d_{i}-\estimate b\right)\right\Vert _{2}+\frac{1}{c}\estimate{\jacobian}_{i,:}\begin{bmatrix}\exact{\location}-\estimate{\ell}\\
s-\estimate b
\end{bmatrix}-\nu_{i}t_{\text{code}}+\beta+\noise_{i}^{(\estimate D,L)}\;.\label{eq:shadow-DL2-1}
\end{equation}
for $\exact{\location}$, $s$, and $\beta$ using in the generalized
least-squares sense, where the Jacobian is evaluated at $\estimate{\location}$
and $t-\estimate d-\estimate b$. 

We can now explain why Van Diggelen's method resolves the integers
only once and iterates only on the continuous unknowns. The $\nu_{i}$'s
that Van Diggelen's method resolves are not equal to the $n_{i}'s$
in the nonlinear Equation~\ref{eq:nonlin-inexact-D}. But when the
linearization error is small enough, the two integer vectors differ
by a constant, $\lfloor\exact b/t_{\text{code}}\rfloor$. This difference
is compensated for by the integer part of the continuous variable
$\beta$, which is not constrained to $[0,t_{\text{code}})$ in the
Gauss-Newton iterations. This is the actual function of shadowing;
to allow $\beta$ to compensate not only for the clock error, but
also for the constant error in $\nu$. When the initial linearization
error is so large that $n-\nu$ is no longer a constant, the method
breaks down.

\section{A Mixed-Integer Least-Squares Approach}

A different approach, which has never been proposed for snapshot positioning,
is to add regularization constraints that will allow us to resolve
all the $4+m$ unknowns in the $m$ instances of Equation~(\ref{eq:linearized-inexact-D-L})
\begin{eqnarray*}
\varphi_{i}t_{\text{code}} & = & \frac{1}{c}\left\Vert \bar{\location}-\controlpoint_{i}\left(t_{i}-d_{i}-\bar{b}\right)\right\Vert _{2}+\frac{1}{c}\jacobian_{i,:}\begin{bmatrix}\exact{\location}-\bar{\ell}\\
\exact b-\bar{b}
\end{bmatrix}-\left(\exact n_{i}-\bar{n}_{i}\right)t_{\text{code}}-\bar{n}_{i}t_{\text{code}}+\left(\exact b-\bar{b}\right)+\bar{b}+\noise_{i}^{(D,L)}
\end{eqnarray*}
using mixed-integer least-squares techniques. Note that we have rewritten
Equation~(\ref{eq:linearized-inexact-D-L}) in a way that emphasizes
a change of variables that facilitate iterative improvements: the
new unknowns are $\exact{\location}-\bar{\ell}$, $\exact b-\bar{b}$,
and $\exact n-\bar{n}$. We initially set $\bar{b}$ and $\bar{n}$
to zero.

We denote the vector of delays by $g$,
\[
g_{i}=\frac{1}{c}\left\Vert \bar{\location}-\controlpoint_{i}\left(t_{i}-d_{i}-\bar{b}\right)\right\Vert _{2}\;.
\]

This section proposes two sets of regularizing equations and explains
how to use this approach in an iterative Gauss-Newton solver. 

\subsection{\label{subsec:apriori-regularization}Resolving the Ambiguities:
Regularization Using A Priori Estimates }

The first set of regularizing equations that we propose are 
\[
\frac{1}{c}\jacobian_{i,:}\begin{bmatrix}\exact{\location}-\bar{\ell}\\
\exact b-\bar{b}
\end{bmatrix}=0\;.
\]
We do not enforce them exactly, only in a (weak) least-squares sense.
They favor solutions of the mixed-integer least-squares problem that
in the vicinity of the a priori solution. This leads to the following
weighted mixed-integer least-squares problem:
\[
\begin{alignedat}{1}\estimate{\location},\estimate b,\estimate n=\arg\min_{\ell,b,n} & \left\Vert W\left(\left[\begin{array}{ll}
\frac{1}{c}\jacobian+\begin{bmatrix}\veczero_{m\times3} & \vecone_{m\times1}\end{bmatrix} & -t_{\text{code}}I_{m\times m}\\
\frac{1}{c}\jacobian & \veczero_{m\times m}
\end{array}\right]\begin{bmatrix}\exact{\location}-\bar{\ell}\\
\exact b-\bar{b}\\
\exact n-\bar{n}
\end{bmatrix}\right.\right.\\
 & \left.\left.-\begin{bmatrix}t_{\text{code}}\varphi-g+\bar{n}t_{\text{code}}-\bar{b}\\
\veczero
\end{bmatrix}\right)\right\Vert _{2}^{2}\;,
\end{alignedat}
\]
where $W$ is a block-diagonal weight matrix derived from the covariance
matrix $C$ of the error terms $\noise$, $W^{T}W=C^{-1}$. Now we
have $2m$ constraints, which for $m\geq4$ should allow us to resolve
the integer $n_{i}$s. 

We propose to choose a diagonal $W$ as follows. We set the first
$m$ diagonal elements of $W$ to the standard deviation of the arrival-time
estimator, say $W_{i,i}=1/\sigma(t_{i})\approx1/(10\,\text{ns})$.
To set the rest, we use box constraints on the a priori estimates
$\bar{\location}$ and $\bar{b}$, denoted
\begin{eqnarray*}
\left|x-\bar{x}\right| & \leq & x_{\max}\\
\left|y-\bar{y}\right| & \leq & y_{\max}\\
\left|z-\bar{z}\right| & \leq & z_{\max}\\
\left|\exact b-\bar{b}\right| & \leq & b_{\max}\;.
\end{eqnarray*}
By the triangle inequality
\[
\left|\jacobian_{i,:}\begin{bmatrix}\exact{\location}-\bar{\ell}\\
\exact b-\bar{b}
\end{bmatrix}\right|\leq\left|\jacobian_{i,1}\right|x_{\max}+\left|\jacobian_{i,2}\right|y_{\max}+\left|\jacobian_{i,3}\right|z_{\max}+\left|\jacobian_{i,4}\right|b_{\max}\;.
\]
We define 
\[
r_{i}=\left|\jacobian_{i,1}\right|x_{\max}+\left|\jacobian_{i,2}\right|y_{\max}+\left|\jacobian_{i,3}\right|z_{\max}+\left|\jacobian_{i,4}\right|b_{\max}\;,
\]
so

\[
\left|\jacobian_{i,:}\begin{bmatrix}\exact{\location}-\bar{\ell}\\
\exact b-\bar{b}
\end{bmatrix}\right|\leq r_{i}\;.
\]
We convert the hard box constraints into soft weighted least squares
in order to allow using a mixed-integer least-squares solver. We need
to set $W_{m+i,m+i}$; if we assume that the error in the constraint
is Gaussian and that an error of $r_{i}/c$ is acceptable (from the
inequality above), then setting  $W_{m+i,m+i}=c/r_{i}$, say, makes
sense. In practice, we use $W_{m+i,m+i}=c/(100\,\text{km})$ in the
experiments below. 

This mixed-integer least squares minimization problem can be solved
by a generic solver, such as one of the solvers that have been developed
for \emph{real-time kinematic }(RTK), a method for solving combined
code-phase and carrier-phase GNSS constraints.

\subsection{\label{subsec:Doppler-Regularization}Doppler Regularization}

It turns out that Doppler shifts allow us to regularize Equation~(\ref{eq:linearized-inexact-D-L})
in a more effective way. GNSS receivers estimate not only the time
of arrival of the signal, but also its Doppler shift. The estimated
Doppler shift is biased, because of the inaccuracy of the receiver's
local (or master) oscillator; it is also inexact. We now show a novel
technique to use the Doppler-shift observations to to regularize Equation~(\ref{eq:linearized-inexact-D-L}). 

Our technique is based on two assumptions. One is that the receiver
is stationary, or more precisely, that its velocity is negligible
relative to the range rate, which is up to about 800~m/s. This assumption
can be easily removed, but its removal leads to additional unknowns
and more complicated expressions that we do not present here. The
other assumption is that the local oscillator and the sampling clock
in the receiver are derived from a single master oscillator in a certain
(very common) way. Again, this assumption can be removed if another
unknown is added.

The Doppler-shift formula for velocities much lower than the speed
of light is
\[
\exact D_{i}\approx-\frac{1}{c}\frac{d}{dt}\left\Vert \exact{\location}-\controlpoint_{i}\right\Vert _{2}f_{0}\;.
\]
The Doppler observations that the receiver makes are
\begin{equation}
D_{i}=-\frac{1}{c}\frac{d}{dt}\left\Vert \exact{\location}-\controlpoint_{i}\right\Vert _{2}f_{0}+\exact f+\epsilon_{i}^{(\delta)}\label{eq:doppler-observation}
\end{equation}
where $\exact f$ is the frequency offset (bias) of the receiver and
$\epsilon_{i}^{(\delta)}$ is an error term that represents the observation
error and the (negligible) slow-speed approximation. Therefore, the
quantities $-cD_{i}/f_{0}$ are biased estimates of the range-rate.
We denote the a priori estimates of the Doppler shifts by $\bar{D}_{i}$. 

We differentiate Equation~(\ref{eq:linearized-inexact-D-L}) by time,
\begin{eqnarray*}
\frac{d}{dt}\left(\varphi_{i}t_{\text{code}}\right) & = & \frac{d}{dt}\left(\frac{1}{c}\left\Vert \bar{\location}-\controlpoint_{i}\left(t_{i}-d_{i}-\bar{b}\right)\right\Vert _{2}+\frac{1}{c}\jacobian_{i,:}\begin{bmatrix}\exact{\location}-\bar{\ell}\\
\exact b-\bar{b}
\end{bmatrix}-n_{i}t_{\text{code}}+\left(\exact b-\bar{b}\right)+\bar{b}\right)+\noise_{i}^{(D,L,\partial)}\;.
\end{eqnarray*}
We first manipulate the equation a bit, to make it easier to differentiate:
\begin{eqnarray}
\frac{d}{dt}\left(c\varphi_{i}t_{\text{code}}-\left\Vert \bar{\location}-\controlpoint_{i}\left(t_{i}-d_{i}-\bar{b}\right)\right\Vert _{2}-c\bar{b}\right) & = & \frac{d}{dt}\left(\jacobian_{i,:}\begin{bmatrix}\exact{\location}-\bar{\ell}\\
\exact b-\bar{b}
\end{bmatrix}-cn_{i}t_{\text{code}}+c\left(\exact b-\bar{b}\right)\right)+\noise_{i}^{(D,L,\partial)}\;.\label{eq:linearized-diff}
\end{eqnarray}
We denote
\[
H=\left[\begin{array}{ccccccccc}
\jacobian_{1,1} & \jacobian_{1,2} & \jacobian_{1,3} & \jacobian_{1,4}+c & -ct_{\text{code}}\\
\vdots & \vdots & \vdots & \vdots &  & \ddots\\
\jacobian_{i,1} & \jacobian_{i,2} & \jacobian_{i,3} & \jacobian_{i,4}+c &  &  & -ct_{\text{code}}\\
\vdots & \vdots & \vdots & \vdots &  &  &  & \ddots\\
\jacobian_{m,1} & \jacobian_{m,2} & \jacobian_{m,3} & \jacobian_{m,4}+c &  &  &  &  & -ct_{\text{code}}
\end{array}\right]\;.
\]
The first three columns of $H$ are identical to those of $\jacobian$,
the next is the fourth column of $\jacobian$ but shifted by $c$,
and the last $m$ columns consist of a scaled identity matrix. We
now express the derivative on the right-hand side of Equation~(\ref{eq:linearized-diff})
as
\[
\frac{d}{dt}\left(H\begin{bmatrix}\exact{\location}-\bar{\ell}\\
\exact b-\bar{b}\\
n_{1}\\
\ldots\\
n_{m}
\end{bmatrix}\right)=H\left(\frac{d}{dt}\begin{bmatrix}\exact{\location}-\bar{\ell}\\
\exact b-\bar{b}\\
n_{1}\\
\ldots\\
n_{m}
\end{bmatrix}\right)+\left(\frac{d}{dt}H\right)\begin{bmatrix}\exact{\location}-\bar{\ell}\\
\exact b-\bar{b}\\
n_{1}\\
\ldots\\
n_{m}
\end{bmatrix}\;.
\]
We assume that the receiver is stationary, so $\exact{\location}-\bar{\ell}$
is time-independent, so $\frac{d}{dt}(\exact{\location}-\bar{\ell})=0$.
The derivatives of the integers $n_{1},\ldots,n_{m}$ are also zero.
The derivative of the remaining element in the vector, $\frac{d}{dt}(\exact b-\bar{b})$,
is not zero and will need to be estimated. It represents the frequency
offset of the receiver, which biases the observed Doppler shift. It
is multiplied by a column whose elements are very close to $c$ (the
range-rate is tiny relative to the speed of light), allowing it to
compensate for the frequency bias.

To differentiate $H$, we exploit the known structure of $\jacobian$.
For each satellite, $\jacobian_{i,1:3}$ is the negation of the so-called
line-of-sight vector $e_{i}^{T}$, which is the normalized direction
from the satellite to the receiver; element $\jacobian_{i,4}$ is
the range-rate. The derivatives of these quantities are shown by Van
Diggelen~\parencite[Equation 8.6]{VanDiggelenAGPS}, Fern\'{a}ndez-Hern\'{a}ndez
and Borre~\parencite{CTNWithoutInitialInfo}, and other sources:
\begin{eqnarray*}
\frac{d}{dt}\jacobian_{i,1:3} & = & \frac{e^{T}\frac{d}{dt}\|\bar{\location}-\controlpoint_{i}\|+\left(\frac{d}{dt}\left(\bar{\location}-\controlpoint_{i}\right)\right)^{T}}{\|\bar{\location}-\controlpoint_{i}\|}\;.
\end{eqnarray*}
where the satellite position $\controlpoint_{i}$ and its velocity
$\frac{d}{dt}\controlpoint_{i}$ are taken at $\hat{t}_{D}$. To reduce
the number of unknowns, we assume that the receiver is stationary,
so $\frac{d}{dt}\bar{\location}=0$, so
\begin{eqnarray*}
\frac{d}{dt}\jacobian_{i,1:3} & = & \frac{e^{T}\frac{d}{dt}\|\bar{\location}-\controlpoint_{i}\|-\left(\frac{d}{dt}\controlpoint_{i}\right)^{T}}{\|\bar{\location}-\controlpoint_{i}\|}\;.
\end{eqnarray*}
Element $\jacobian_{i,4}$ is the range-rate of satellite $i$, so
its derivative with respect to time is the satellite's range acceleration,
\begin{eqnarray*}
\frac{d}{dt}\jacobian_{i,4} & =\frac{d^{2}}{dt^{2}} & \|\bar{\location}-\controlpoint_{i}\|\;.
\end{eqnarray*}
We use finite differences to evaluate this second derivative. The
fourth column of $H$ is $\jacobian_{:,4}+c$, but the derivative
of $c$ is obviously zero. The derivative of $-ct_{\text{code}}$
is also zero, so
\[
\left(\frac{d}{dt}H\right)\begin{bmatrix}\exact{\location}-\bar{\ell}\\
\exact b-\bar{b}\\
n_{1}\\
\ldots\\
n_{m}
\end{bmatrix}=\left(\frac{d}{dt}H_{:,1:4}\right)\begin{bmatrix}\exact{\location}-\bar{\ell}\\
\exact b-\bar{b}
\end{bmatrix}\;.
\]
We now derive the left-hand side of Equation~(\ref{eq:linearized-diff}),
\[
\frac{d}{dt}\left(c\varphi_{i}t_{\text{code}}-\left\Vert \bar{\location}-\controlpoint_{i}\left(t_{i}-d_{i}-\bar{b}\right)\right\Vert _{2}-c\bar{b}\right)\;.
\]
The derivative of the a prioi range estimate $\left\Vert \bar{\location}-\controlpoint_{i}\left(t_{i}-d_{i}-\bar{b}\right)\right\Vert _{2}$
is the a priori range-rate, which we can compute. The derivative of
$c\bar{b}$ is zero. 

To understand the first term, recall that 
\[
\left(n_{i}+\varphi_{i}\right)t_{\text{code}}=\frac{1}{c}\left\Vert \exact{\location}-\controlpoint_{i}\left(t_{D,i}\right)\right\Vert _{2}+\exact b+\noise_{i}\;.
\]
so
\[
c\frac{d}{dt}\varphi_{i}t_{\text{code}}=\frac{d}{dt}\left\Vert \exact{\location}-\controlpoint_{i}\left(t_{D,i}\right)\right\Vert _{2}+c\frac{d}{dt}\exact b+c\frac{d}{dt}\noise_{i}\;.
\]
We now rewrite (\ref{eq:doppler-observation}) as
\[
\frac{1}{c}\frac{d}{dt}\left\Vert \exact{\location}-\controlpoint_{i}\right\Vert _{2}f_{0}=-D_{i}+\exact f+\epsilon_{i}^{(\delta)}
\]
or
\[
\frac{d}{dt}\left\Vert \exact{\location}-\controlpoint_{i}\right\Vert _{2}=-\frac{cD_{i}}{f_{0}}+\frac{c\exact f}{f_{0}}+\frac{c}{f_{0}}\epsilon_{i}^{(\delta)}\;.
\]
We now substitute in the left-hand side of Equation~(\ref{eq:linearized-diff}):
\begin{eqnarray*}
\frac{d}{dt}c\varphi_{i}t_{\text{code}} & = & -\frac{cD_{i}}{f_{0}}+\frac{c\exact f}{f_{0}}+\frac{c}{f_{0}}\epsilon_{i}^{(\delta)}+c\frac{d}{dt}\exact b+c\frac{d}{dt}\noise_{i}\;.
\end{eqnarray*}
The term $\exact f/f_{0}$ is the relative local-oscillator error
in the receiver. If the oscillator runs too fast, $\exact f$ is negative.
Assuming that all the clocks in the receiver are derived from a master
oscillator, if it runs too fast, $\exact b$ grows over time. Under
this assumption
\[
-\frac{c\exact f}{f_{0}}=c\frac{d}{dt}\exact b
\]
so these terms cancel each other. If our assumption on the receiver
does not hold, we would need to estimate 
\[
\frac{c\exact f}{f_{0}}+c\frac{d}{dt}\exact b\;.
\]

That's it. We have arrived at a system of $m$ linear equations that
we use to regularize the mixed-integer equations. The equations are:
\begin{equation}
-\frac{c}{f_{0}}D-\frac{d}{dt}\left\Vert \exact{\location}-\controlpoint\right\Vert _{2}=H_{:,4}\left(\exact u-\bar{u}\right)+\left(\frac{d}{dt}H_{:,1:4}\right)\begin{bmatrix}\exact{\location}-\bar{\ell}\\
\exact b-\bar{b}
\end{bmatrix}\;.\label{eq:doppler-regularization}
\end{equation}
In this equation, $D$ represents the vector of observed Doppler shifts,
$\frac{d}{dt}\left\Vert \exact{\location}-\controlpoint\right\Vert _{2}$
is the vector of the a priori range rates, and $\exact u-\bar{u}=\frac{d}{dt}\left(\exact b-\bar{b}\right)$
is a new scalar unknown. We have explained above how to compute $H_{:,4}$
and $\frac{d}{dt}H_{:,1:4}$. The full regularized weighted least-squares
that we solve is
\[
\begin{alignedat}{1}\estimate{\location},\estimate b,\estimate n,\estimate u=\arg\min_{\ell,b,n,u} & \left\Vert W\left(\left[\begin{array}{llc}
\frac{1}{c}\jacobian+\begin{bmatrix}\veczero_{m\times3} & \vecone_{m\times1}\end{bmatrix} & -t_{\text{code}}I_{m\times m} & \veczero_{m\times1}\\
\left(\frac{d}{dt}H_{:,1:4}\right) & \veczero_{m\times m} & H_{:,4}
\end{array}\right]\begin{bmatrix}\exact{\location}-\bar{\ell}\\
\exact b-\bar{b}\\
\exact n-\bar{n}\\
\exact u-\bar{u}
\end{bmatrix}\right.\right.\\
 & \left.\left.-\begin{bmatrix}t_{\text{code}}\varphi-g+\bar{n}t_{\text{code}}-\bar{b}\\
-\frac{c}{f_{0}}D-\frac{d}{dt}\left\Vert \exact{\location}-\controlpoint\right\Vert _{2}
\end{bmatrix}\right)\right\Vert _{2}^{2}\;.
\end{alignedat}
\]

\subsection{Iterating to Cope with Large A Priori Errors}

Solving the linearized and regularized mixed-integer least-squares
problem improves the initial a priori estimates of $\exact{\location}$
and $\exact b$, but not to the extent possible given the code phases.
The most important factor that limits the accuracy of the corrections
is the fact that when the a priori estimates are large, the resolved
integers, the $n_{i}$'s, are inexact. Therefore, we incorporate the
mixed-integer solver into a Gauss-Newton-like iteration in which we
correct all the unknowns, including the integer ambiguities, more
than once. 

More specifically, once we solve the mixed-integer least-squares problem
for $\exact n-\bar{n}$, $\exact{\location}-\bar{\ell}$ and $\exact b-\bar{b}$
(and for $\exact u-\bar{u}$ in the Doppler formulation), we use the
corrections to improve the estimates of the receiver's location and
of the departure times and we linearize Equation~(\ref{eq:nonlin-inexact-D})
again. We now solve the newly-linearized least-squares problem again
for additional corrections, and so on.

\section{\label{sec:Implementation-and-Evaluation}Implementation and Evaluation}

We have implemented all the methods that we described above in MATLAB. 

We use Borre's Easy Suite~\parencite{EasySuite,EasySuite2} to perform
many routine calculations. In particular, we use it to correct GPS
time (check\_t), to correct for Earth rotation during signal propagation
time (e\_r\_corr), to read an ephemeris from a RINEX file and to extract
the data for a particular satellite (rinexe, get\_eph and find\_eph),
to transform Julian dates to GPS time (gps\_time), to represent Julian
dates as one number (julday), to compute the coordinate of a satellite
at a given time in ECEF coordinates (satpos), to compute the azimuth,
elevation, and distance to a satellite (topocent, which calls togeod
to transform ECEF to WG84 coordinates), and to approximate the tropospheric
delay (tropo). We also use a MATLAB function by Eric Ogier (ionophericDelay.m,
available on the MathWorks File Exchange) to approximate the ionopheric
delay using the Klobuchar model. We take the parameters for the Klobuchar
model from files published by the GNSS Research Center at Curtin University\footnote{\texttt{http://saegnss2.curtin.edu/ldc/rinex/daily/}}. 

During the Gauss--Newton phase of the algorithm (after the integers
have been determined), if we have only 4 observations, we add a pseudo-measurement
constraint that constrains the correction to maintain the height of
the target, in the least-squares sense~\parencite{VanDiggelenAGPS}.

We use Chang and Zhou's MILES package~\parencite{MILES} to solve mixed
integer least-squares problems. 

We take ephemeris data from RINEX navigation files published by NASA\footnote{\texttt{https://cddis.nasa.gov/archive/gnss/data/daily/}}. 

We filter satellites that are lower than 10 degrees above the horizon
which have the lowest SNR and are more likely to suffer from multipath
interference.

We evaluated the code on data from several sources:
\begin{itemize}
\item Publicly available observation data files in a standard format (RINEX)
distributed by NASA. We used these to test our algorithms in the initial
phases of the research. These results are not shown here.
\item GPS Simulations. We generated satellite positions ephemeris files
and used them to compute times of arrival and code phases. These simulations
do not include ionospheric or tropospheric delays, so they help us
separate the issues arising from these delays from other algorithmic
issues.
\item Code-phase and Doppler-shift measurements collected by us using a
u-blox ZED-F9P GNSS receiver, connected to an ANN-MB-00 u-blox antenna
mounted on a steel plate on top of a roof with excellent sky view.
We established the precise coordinates of the antenna (to compute
errors) using differential carrier-phase corrections from a commercial
virtual reference station (VRS)\footnote{\texttt{https://axis-gps.com}}.
The WGS84 coordinates of the antenna are 32.1121756, 34.8055775 with
height above sea level of 61.15~m. The code phase measurements are
included in the UBX-RXM-MEASX emitted by the receiver. The data set
includes about 700 epochs, one every minute (so they span a little
more than 11 hours). The number of satellites per epoch ranges from
8 to 13 and after filtering by elevation, between 7 and 11.
\item Recordings of RF samples made by a bat-tracking GPS snapshot logger.
The tag model is called Vesper. It was designed and produced by Alex
Schwartz Developments on the basis of an earlier tag called Robin
designed and produced by a company called CellGuide that no longer
exists. The tag records 1-bit RF samples at a rate of 1023000 samples
per second. (The sampling rate is a multiple of $1/t_{\text{code}}$;
this is known to make time-of-arrival estimate difficult~\parencite{GNSSSamplingRates2018}
but the rate cannot be changed in this logger). The tag was configured
to record a 256~ms sample every 10 minutes for a few hours. It was
placed next to the ANN-MB-00 antenna. 
\end{itemize}
\begin{figure}
\hfill{}\includegraphics[width=1\textwidth]{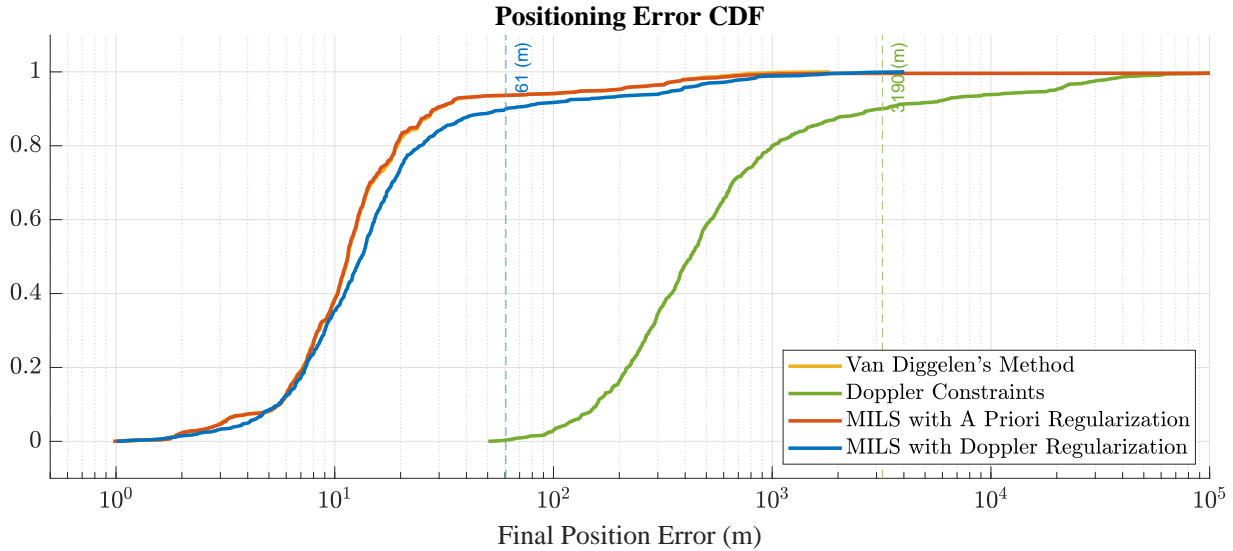}\hfill{}

\caption{\label{fig:overall-cdf}Cumulative distribution function of the absolute
positioning errors of four algorithms: Van Diggelen's non-iterrative
method, the Doppler constraints alone (the first phase of Fern\'{a}ndez-Hern\'{a}ndez
and Borre's method), and MILS with either a priori or Doppler regularization.}
\end{figure}

Figure~\ref{fig:overall-cdf} shows the cumulative distribution function
of four algorithms: Van Diggelen's non-iterative method, the Doppler
constraints alone (as used in the first phase of Fern\'{a}ndez-Hern\'{a}ndez
and Borre's method), and MILS with either a priori or Doppler regularization.
The data from the u-blox receiver was used to produce these graphs.
We used all 694 epochs. The initial error was of 20--21~s (uniform
distribution) and 20~km in a random uniform horizontal direction.
In the MILS algorithms, the final position was computed with the regularization
constraints; this is why the Doppler regularization produced less
accurate results. We can see that the accuracy of MILS with a priori
regularization and of Van Diggelen's method are essentially identical.

\begin{figure}
\includegraphics[width=0.47\textwidth]{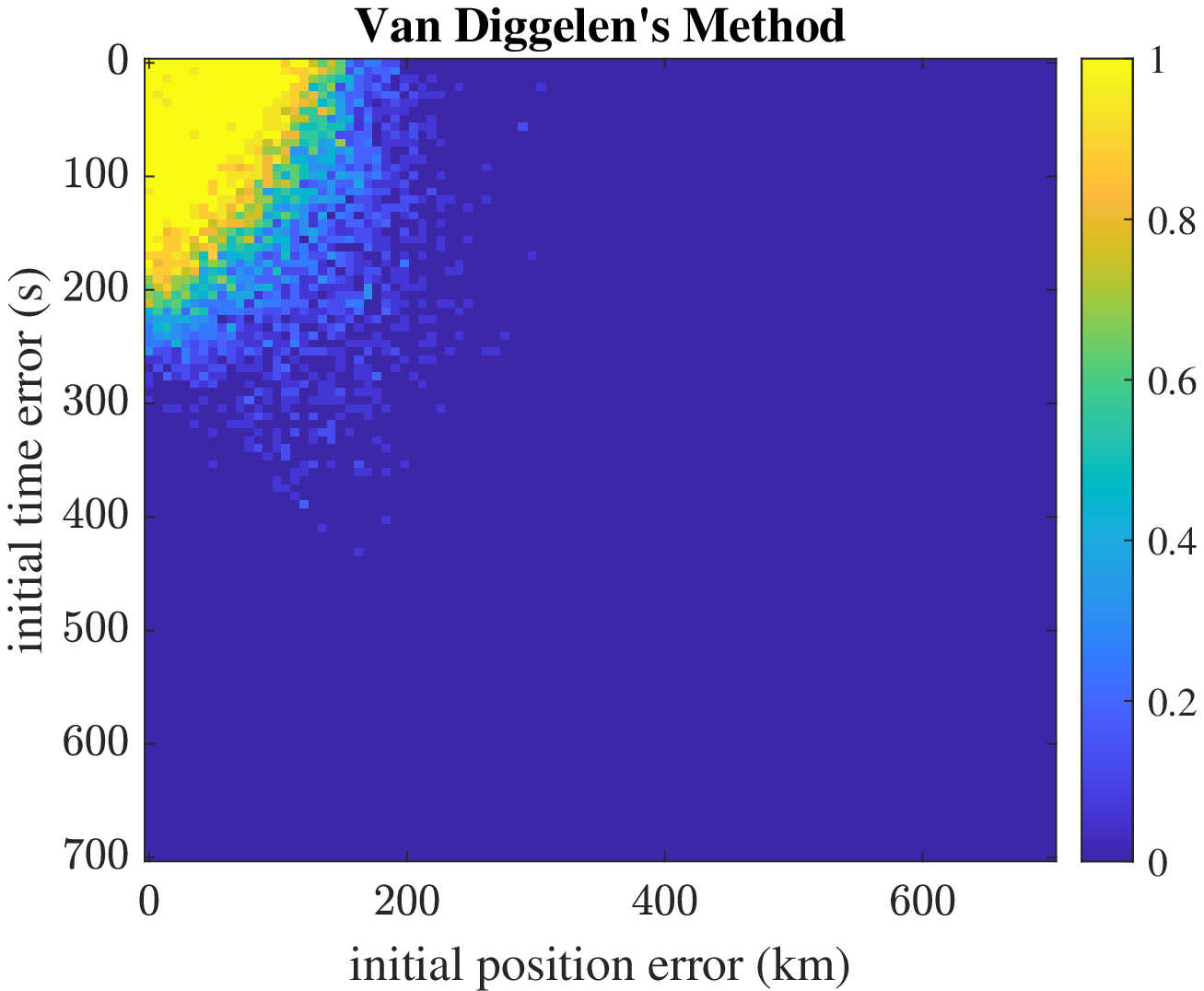}\hfill{}\includegraphics[width=0.47\textwidth]{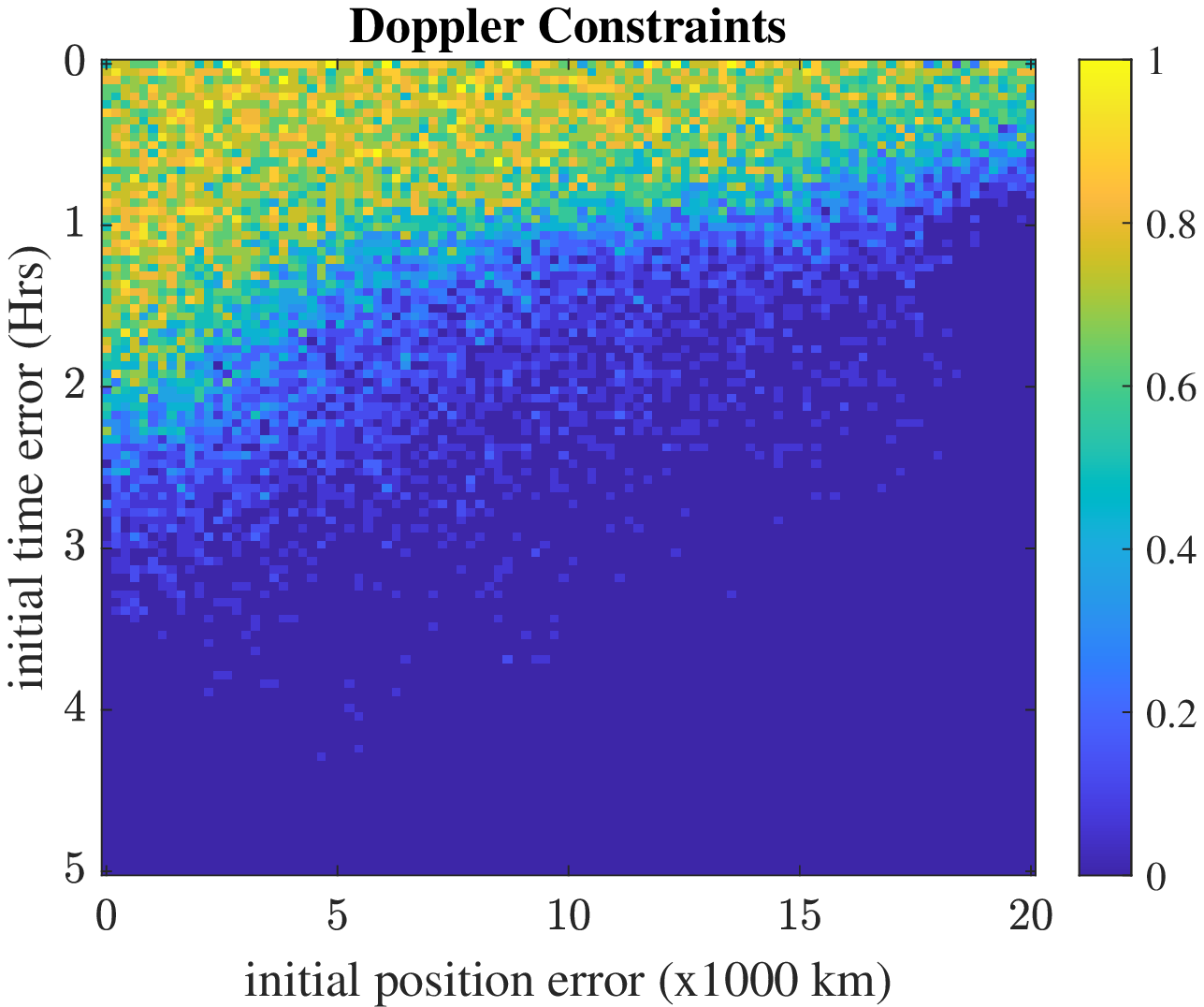}\\
~

\includegraphics[width=0.47\textwidth]{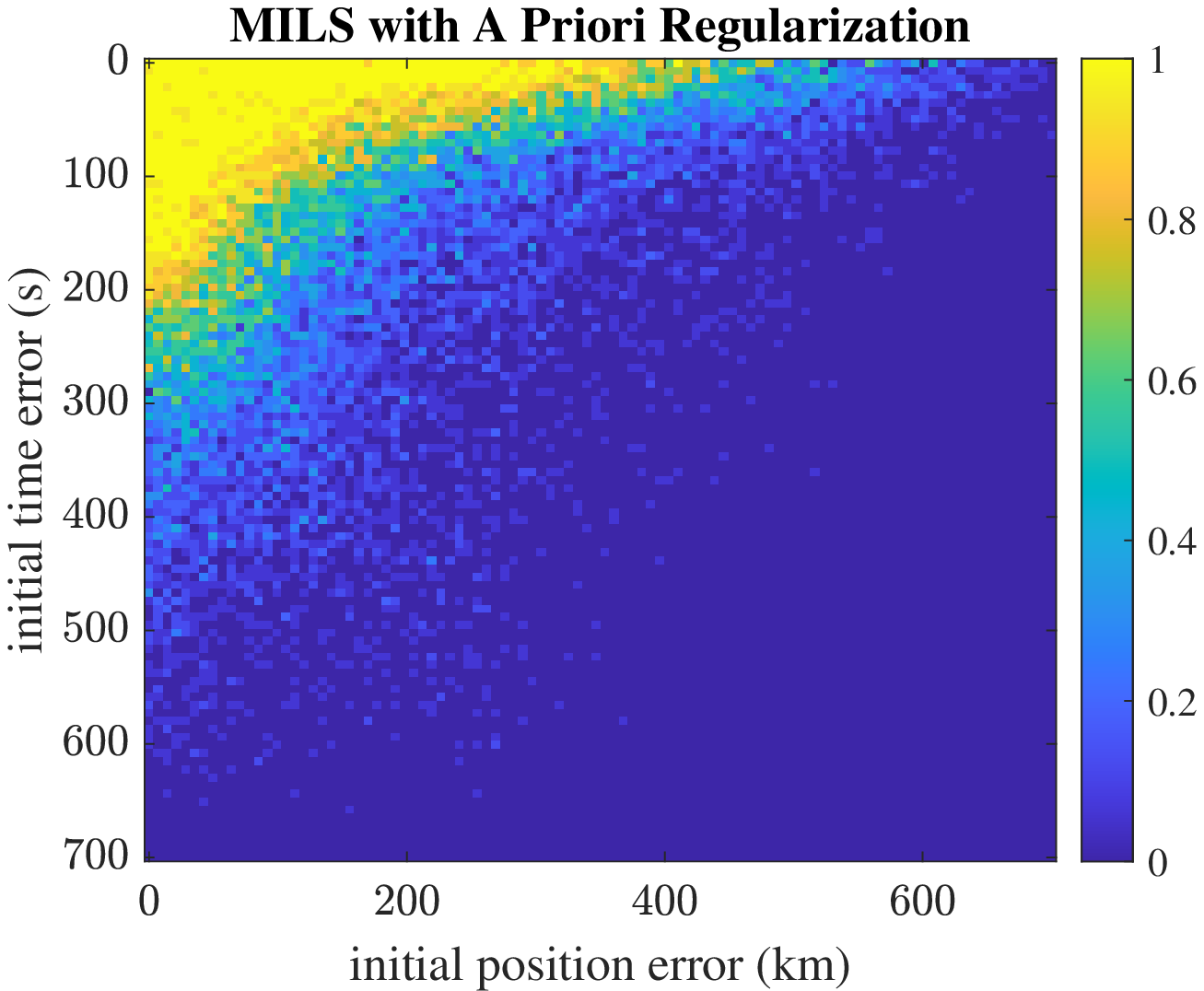}\hfill{}\includegraphics[width=0.47\textwidth]{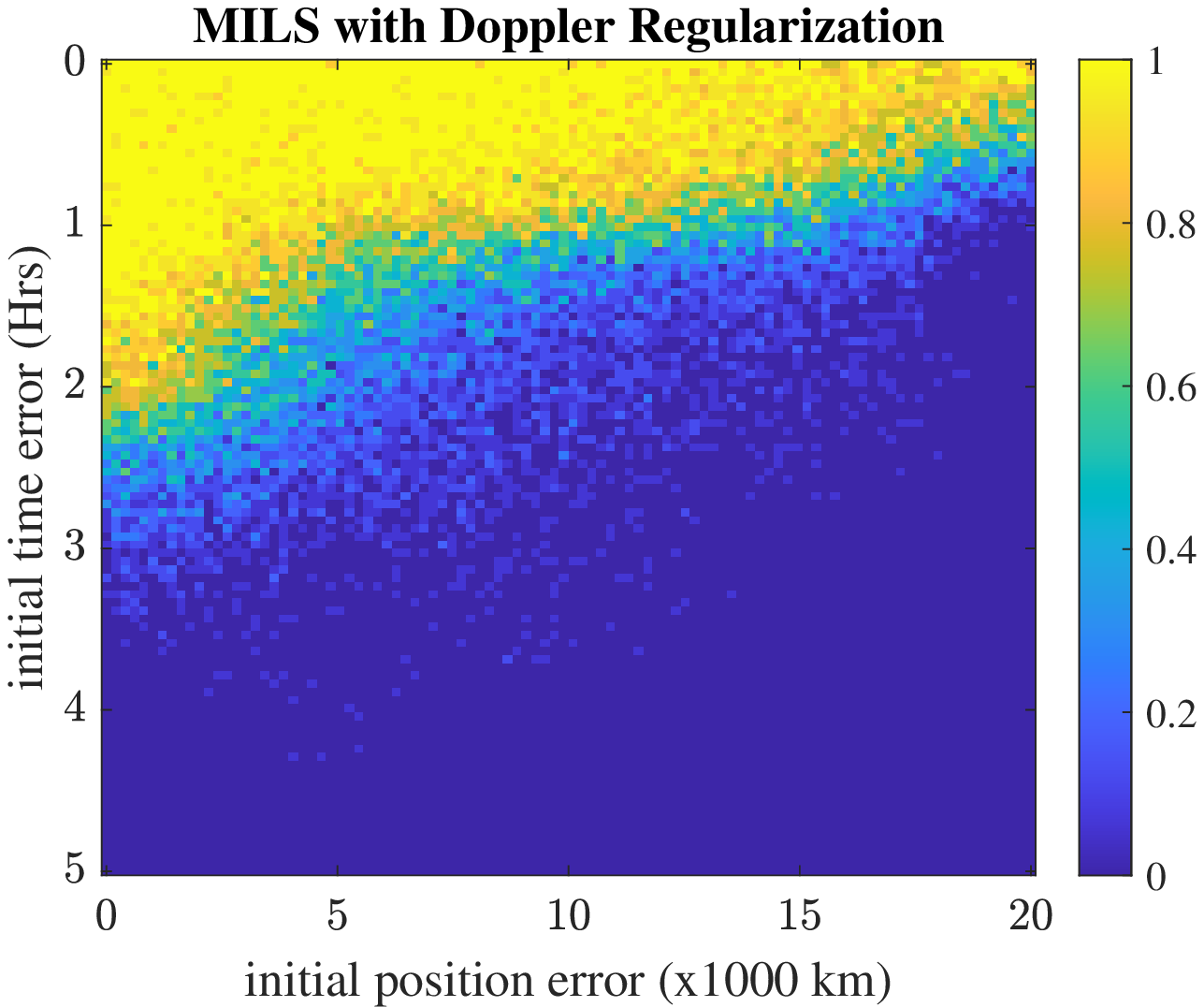}

\caption{\label{fig:success-heatmaps}The probability of obtaining a fix with
an error smaller than 1~km from the u-blox data set using four different
algorithms.}
\end{figure}
Figure~\ref{fig:success-heatmaps} compares the probability of success
achieved by our regularized mixed-integer least-squares (MILS) solver
with that achieved by Van Diggelen's non-iterative method and by the
Doppler constraints alone. We considered fixes that are within 1~km
of the true location to be a success in obtaining the correct integer
values. Each pixel in these heat maps represents 16 different runs.
Each run uses a random epoch, a random initial location estimate,
and a random initial time estimate. The initial location estimates
have a given distance to the true location (the $x$ axis of the heat
map) but a random azimuth. The initial time estimate is a slight perturbation
(uniform between zero and one second) of the given time error, which
is the $y$ axis of the heat map. In each pixel, half of the initial
time errors are positive and half are negative. We used all the satellites
in view in each epoch. 

The results clearly show that the MILS algorithm, even with the simple
a priori time and location regularization from Section~\ref{subsec:apriori-regularization},
outperforms Van Diggelen's non-iterative method. Van Diggelen's method
obtains a correct fix in almost all cases (success probability close
to 1) when the initial location error is small and the initial time
error is 150~s or less, when the initial location time error is small
and the initial location error is 100~km or less, and in other equivalent
combination of time and location errors. The corresponding limits
for the MILS algorithm with a priori regularization are about 150~s
and 250~km. 

Doppler-shift observations expand dramatically the region of convergence
in both approaches. The MILS algorithm with Doppler regularization
obtains a correct fix as long as the initial time error is at up to
about 180~s (3 minutes); this works even with great-circle distances
of 20,000~km, which means that the initial position can be essentially
anywhere on Earth. If the initial position error is small, the method
can tolerate initial time errors of up to about 80 minutes (5000~s).
The heat map of the Doppler constraints alone, together with the CDF
in Figure~\ref{fig:overall-cdf}, indicate that these constraints
produce an estimate good enough for initializing Van Diggelen's method,
but are not accurate enough on their own. Indeed, Van Diggelen writes
about the Doppler constraints alone: ``For less than 1~Hz of measurement
error, we expect a position error of the order of 1~km''~\parencite[ Section~8.3]{VanDiggelenAGPS};
this explains why the probabilities in the top-right plot in Figure~\ref{fig:success-heatmaps}
are usually far from~1, even with small initial errors. In general,
both approaches have similar regions of convergence and they produce
similarly-accurate fixes.

\begin{figure}
\includegraphics[width=0.47\textwidth]{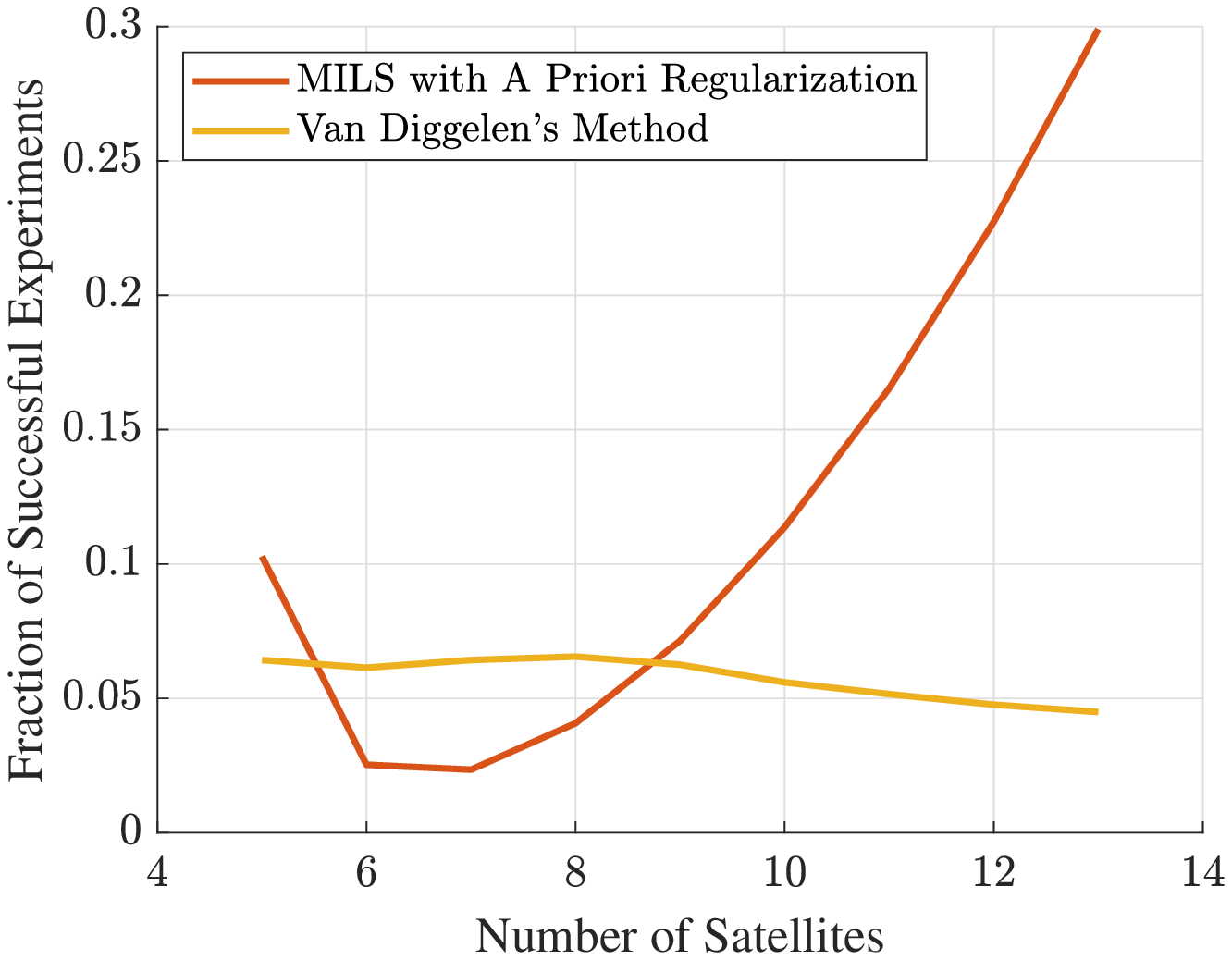}\hfill{}\includegraphics[width=0.47\textwidth]{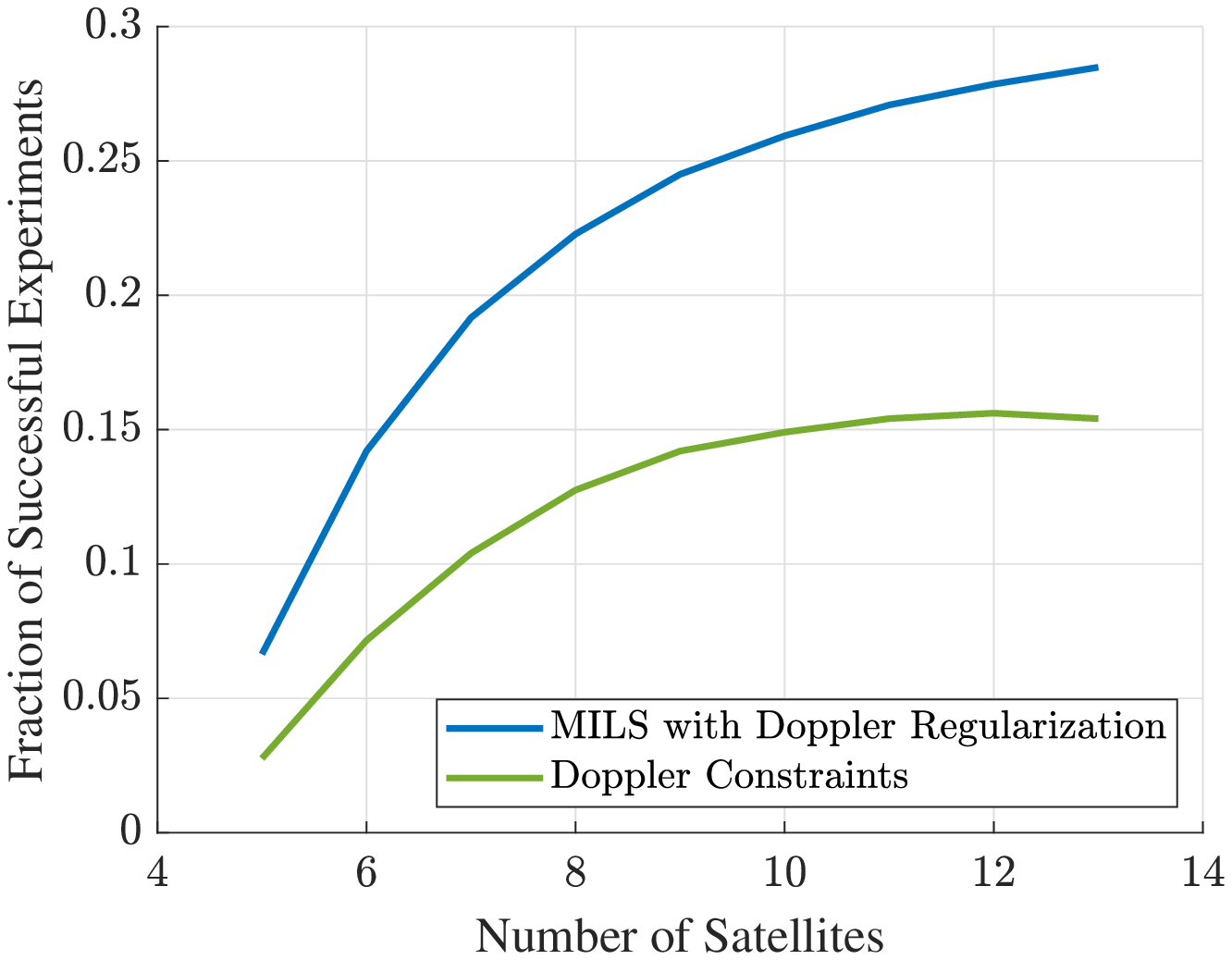}

\caption{\label{fig:nsats}The fraction of successful positioning (error of
at most 1~km) in the spaces of initial errors shown in Figure~\ref{fig:success-heatmaps}
as a function of the number of satellites (observations) used.}
\end{figure}
Figure~\ref{fig:nsats} explores now the number of satellites (observations)
affects the success rates of the four methods. We repeated the experiment
whose results are shown in the heat maps in Figure~\ref{fig:success-heatmaps},
but only on the 20 epochs in which 13 satellites were in view. We
selected random subsets of the satellites in view and random initial
errors, within the bounds shown in Figure~\ref{fig:success-heatmaps},
and computed the fraction of successful experiments. We can see that
when only code phases are used, Van Diggelen's method is better when
using 6--8 observations, probably because the weighting of the observations
in the MILS method sometimes leads to incorrect integers when Van
Diggelen's method resolves the integers correctly. However, with 5
satellites in view or more than 8, the MILS method is better. MILS
with Doppler regularization is superior to all the other methods.

While our Matlab implementation is not designed to carefully evaluate
running times and computational efficiency, we did measure the running
times and we can draw from them some useful conclusions. The running
times of a single Gauss-Newton correction step in Van Diggelen's method
and in the solution of the Doppler equations is 8--10~$\mu\text{s}$,
while the running time of a single Gauss-Newton correction step in
the MILS formulation is about 2.5~ms when using a priori regularization
and 0.6~ms when using Doppler regularization. While the MILS methods
are clearly more expensive, they also appear to be fast enough for
real-time applications.

\section{\label{sec:Conclusions-and-Discussion}Conclusions and Discussion}

We have shown that Van Diggelen's ingenious \emph{coarse-time navigation}
algorithm~\parencite{IEEEexample:uspat,VanDiggelenAGPS} that estimates
a location from GNSS observations without departure times is essentially
a specialized solver for a mixed-integer least-squares problem. Even
though Van Diggelen's algorithms has been cited and used by many authors,
the actual form of the mixed-integer optimization problem has never
been presented; we present it in this paper for the first time.

We also show that the integer ambiguities can be resolved by regularizing
the mixed-integer least-squares problem. We proposed two regularization
techniques, one that biases solutions towards an initial a priori
estimate. This extends Van Diggelen's use of the a priori estimate
to resolve the integers, but our regularization approach can resolve
the integers with larger initial errors than Van Diggelen's. In effect,
the general mixed-integer formulation uses the available information
more effectively than Van Diggelen's specialized solver. 

We also proposed a regularization method based on Doppler-shift observations.
This method allows our solver to resolve the correct integers even
with huge initial time or position errors. Doppler shifts have been
used in snapshot positioning before, but they were always used to
produce an initial position and time estimate that is subsequently
used as an a priori estimate in Van Diggelen's algorithm. This approach,
due to Fern\'{a}ndez-Hern\'{a}ndez and Borre~\parencite{CTNWithoutInitialInfo},
is also extremely effective.

Our algorithm iterates over the entire mixed-integer least-squares
problem more than once. If one resolves the integers once, in the
first iteration, and continues to iterate only on the continuous unknowns,
using the resolved integers, the method converges, but to fixes with
larger errors.

In effect, by cleanly formulating the mixed-integer optimization problem
that underlies snapshot positioning, we enabled the exploration of
a wide range of solvers, including the two regularized solvers that
we presented here. We believe that additional solvers can be discovered
for this formulation. In contrast, all prior research treated Van
Diggelen's algorithm as a clever black box, limiting the range of
algorithms that can be developed.

Our new methods are inspired by the \emph{real-time kinematic} (RTK)
method, which resolves a position from both code-phase and carrier-phase
GNSS observations~\parencite{GNSSHandbookIntegerAmbiguity}. In RTK, the
position is eventually resolved by carrier-phase constraints, which
have integer ambiguities; these constraints are regularized by pseudo-range
constraints, which are less precise but have no integer ambiguities.
Here the position is resolved by integer-ambiguous code-phase constraints,
which are regularized by either a priori estimates or by Doppler-shift
observations. RTK also requires so-called differential constraints
at a fixed receiver, because the carrier phase of the satellites are
not locked to each other. Here we do not require differential corrections
because the code departure times are locked to a whole milliseconds
in all the satellites.

\paragraph*{Acknowledgments}

This study was also supported by grant 1919/19 from the Israel Science
Foundation. Thanks to Aya Goldshtein for assisting with collecting
data from the bat-tracking GPS snapshot logger. Thanks to Amir Beck
for helpful discussions. Thanks to the three reviewers for comments
and suggestions that helped us improve the paper.

\printbibliography

\end{document}